\documentclass[aps,prl,reprint,twocolumn,amsmath,amssymb,citeautoscript]{revtex4-1}
\usepackage{graphicx}
\usepackage{dcolumn}
\usepackage{bm}
\usepackage{latexsym,epsfig}
\usepackage{graphicx}
\usepackage{verbatim}
\usepackage{comment}
\usepackage{amsmath}
\usepackage{amssymb}
\usepackage{stmaryrd}
\usepackage{color}
\usepackage{epstopdf}
\usepackage{grffile}
\usepackage{lipsum}
\usepackage{enumitem}
\DeclareGraphicsExtensions{.eps}

\newcommand{\beq}{\begin{equation}}
\newcommand{\eeq}{\end{equation}}
\newcommand{\bea}{\begin{eqnarray}}
\newcommand{\eea}{\end{eqnarray}}
\newcommand{\ben}{\begin{eqnarray*}}
\newcommand{\een}{\end{eqnarray*}}
\newcommand{\bfig}{\begin{figure}}
\newcommand{\efig}{\end{figure}}

\usepackage{hyperref}
\hypersetup{
    colorlinks=true,      
    urlcolor=blue,
    citecolor=blue,
    linkcolor=blue
}

\begin{document}
\title{Emergence of multiple localization transitions in a one-dimensional quasiperiodic lattice }
\author{Ashirbad Padhan$^1$, Mrinal Kanti Giri$^1$, Suman Mondal$^1$ and Tapan Mishra$^{1,2,3}$}
\email{mishratapan@gmail.com}
\affiliation{$^1$Department of Physics, Indian Institute of Technology, Guwahati-781039, Assam, India}
\affiliation{$^2$National Institute of Science Education and Research, Jatni, Odisha 752050, India}
\affiliation{$^3$Homi Bhabha National Institute, Training School Complex, Anushakti Nagar, Mumbai 400094, India.}


\date{\today}

\begin{abstract}
Low dimensional quasiperiodic systems exhibit localization transitions by turning all quantum states localized after a critical quasidisorder. While certain systems with modified or constrained quasiperiodic potential undergo multiple localization transitions in one dimension, we predict an emergence of multiple localization transitions without directly imposing any constraints on the quasiperiodic potential. By considering a one-dimensional system described by the Aubry-Andr\'{e} (AA) model, we show that an additional staggered onsite potential can drive the system through a series of localization transitions as a function of the staggered potential. Interestingly, we find that the number of localization transitions strongly depends on the strength of the quasiperiodic potential. 
Moreover, we obtain the signatures of these localization transitions in the expansion dynamics and propose an experimental scheme for their detection in the quantum gas experiment.

\end{abstract}





\maketitle
{\em Introduction.-} 
The quasiperiodic lattices offer unique opportunities to study localization transitions compared to the random lattices. While for the systems with random disorder, an arbitrarily small disorder localizes all the single particle states in low dimensions~\cite{gang_of_four_PRL79,Lee}, the quasiperiodic lattices exhibit localization transition through a critical quasidisorder~\cite{paredesreview,review2}.  Due to the much easier accessibility of the quasiperiodic lattices over the random lattices,  localization transitions have been observed in various recent experiments in the context of the AA model~\cite{roati,lueschen,schreiber}.

One of the simplest but widely explored quasiperiodic models which exhibits localization transition is the Aubry-Andr\'{e} (AA) model~\cite{AA}. Owing to the self-dual nature, the one-dimensional (1D) AA model exhibits a sharp localization transition at a critical quasidisorder turning all extended single particle eigenstates localized~\cite{AA,Jitomirskaya}. However, breaking the self-duality of the AA model leads to a localization transition through an intermediate phase  with coexisting extended and localized eigenstates separated by a critical energy known as the mobility edge (ME) - a situation similar to Anderson localization in three-dimensions~\cite{Mott_1987,anderson}. Numerous theoretical investigations have predicted the existence of the intermediate phases and the ME in 1D quasiperiodic lattices~\cite{biddle, Ganeshan, dassarma2017,dassarma2020, subrotoreview, ray_2015,Boers,shuchen_mosaic} leading to their successful experimental observations in recent years~\cite{lueschen,An,Gadway2021,lado2020_nphys}.

In general, it is understood that after the localization transition in quasiperiodic systems, the localized states remain localized as a function of the quasidisorder (hereafter referred as only disorder). However, a recent study by some of us has revealed the phenomenon of  the re-entrant localization transition by imposing certain constraints on the parameters of the AA model ~\cite{mishrareentrant}. It has been shown that a dimerized hopping strength along with staggered onsite disorder, drives the system through two localization transitions. In other words, after a first localization transition, some of the localized states become extended for a range of parameters and eventually become localized at a different critical disorder strength. On the other hand, a cascade of localization transitions has been predicted and observed~\cite{lado2020_nphys} in a system with an engineered quasiperiodic potential that continuously maps the AA model to the Fibonacci model~\cite{kohmoto,ostlund}. These remarkable findings further motivate to explore the possible existence of such multiple localization transitions in a simple quasiperiodic model without directly imposing any constraint on the quasidisorder.

In this letter, we show that in a 1D quasiperiodic lattice described by the AA model, an additional staggered onsite potential can drive the system through multiple localization transitions. We show that for fixed disorder, the system undergoes a series of localization transitions as a function of the staggered potential. Interestingly, the model allows to control the number of localization transitions by appropriately tuning the disorder strength. We further obtain the signatures of these multiple localization transitions in the experiments on expansion dynamics and provide an experimental scheme for their observation using ultracold atoms in optical lattices.

{\em Model.-} The one-dimensional quasiperiodic lattice model which is considered for our studies is given by 
\begin{eqnarray}
 \mathcal{H}=&-&J\sum_i(c_i^\dagger c_{i+1}+H.c.) +\lambda\sum_i \cos(2\pi\beta i+\phi)n_i \nonumber \\
 &+&\Delta\sum_i(-1)^in_i
\label{eq:ham}
\end{eqnarray} 
where $c_{i}^{\dagger}$ and $n_i=c_{i}^{\dagger} c_{i}$ are the fermionic creation and the number operator respectively at site $i$. $J$ is the hopping matrix element between the nearest-neighbour sites and  $\lambda$ corresponds to the disorder strength. Unless mentioned otherwise, all the numerical simulations are performed under open boundary condition (OBC). The quasiperiodicity is ensured by choosing irrational numbers $\beta=(\sqrt{5}-1)/2$ - known as the inverse Golden ratio and $\beta=F_{n-1}/F_n$ with $F_n$ being the $n$-th Fibonacci number for open and periodic boundary condition (PBC) respectively. The ratios of Fibonacci numbers are used to protect the quasiperiodicity in the PBC~\cite{paredesreview, Modugno_2010, dassarma2020, gaa} which tend to the inverse Golden ratio for large system sizes.
$\phi$ is the phase offset between the primary and secondary lattices and $\Delta$ is the onsite staggered potential. We set $J=1$ as the unit of energies and considered system sizes according to the Fibonacci series. Note that when $\Delta=0$, the Eq.~\ref{eq:ham} represents the standard AA model which exhibits a sharp localization transition at $\lambda_{AA}=2$.  For our analysis we consider large system sizes such that the effect of $\phi$ is negligible. Therefore, for most of the cases we set $\phi=0$ unless otherwise mentioned. It is important to note that the lattice corresponding to Eq.~\ref{eq:ham} has already been used as a tool to observe the localization transitions~\cite{lueschen,schreiber, bordia}. However, the localization transition of the model corresponding to Eq.~\ref{eq:ham} itself has not been explored in detail except for some specific parameter domain where a re-entrant localization of individual states has been predicted~\cite{sil}. In the following we explore the physics of this simple yet not well explored model and predict the phenomenon of multiple localization transitions. 

\begin{figure}[t]
\centering
{\includegraphics[width=3.4in]{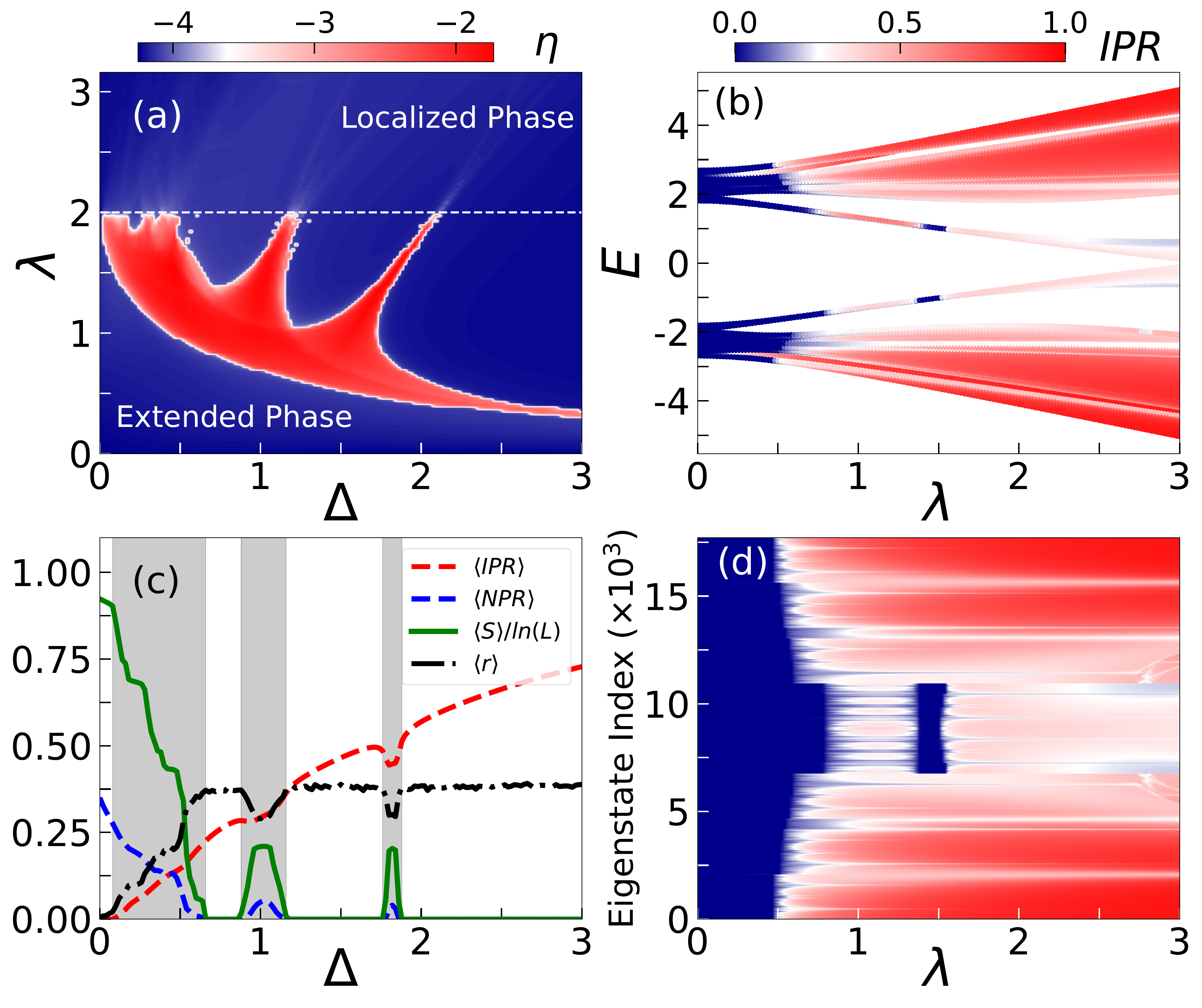}}
\caption{(a) The phase diagram in the $\Delta$ - $\lambda$ plane obtained using the values of $\eta$ for a system of size $L=17711$.  (b) IPR of all the eigenstates as a function of energies and $\lambda$ for $\Delta=1.8$ and a system size of $L=17711$. (c) Extrapolated values of $\langle \rm IPR\rangle$ (dashed red), $\langle \rm NPR\rangle$ (dashed blue) with system sizes $L=2584,~4181,~6765,~10946,~17711$, $\langle S\rangle/\ln(L)$ (solid green) for $L=17711$ and $\langle r \rangle$ (dot-dashed black) for $L=2584$ are plotted as a function of $\Delta$ for $\lambda=1.5$ showing the multiple localization transitions. The $\langle r \rangle$ is computed using PBC with $\beta=1597/2584$ and an average over $5000$ phase offsets $\phi$ are considered.} The intermediate phases are indicated by the grey shaded regions. (d) IPR of all the eigenstates as a function of state index and $\lambda$ for $\Delta=1.8$ and $L=17711$. The white dashed line in (a) indicates the AA critical  point at $\lambda=2$.
\label{fig:fig1}
\end{figure}

{\em Multiple localization transitions.-} In this section we study the localization transition of the model given in Eq.~\ref{eq:ham} and the central result is given as the phase diagram in the $\Delta$ - $\lambda$ plane as depicted in Fig.~\ref{fig:fig1}(a). The blue region below (above) the red region is the extended (localized) phase and the red region corresponds to the intermediate phase where both extended and localized states exist. It can be seen that the system undergoes localization transitions (from all states extended to all states localized) through the intermediate region as a function of $\lambda$ for all values of $\Delta$ except at $\Delta=0$ (the AA limit). This is expected as the self-duality of Eq.~\ref{eq:ham} is broken due to the onset of $\Delta$. However, for some particular values of $\Delta$, two localization transitions or the re-entrant localization transitions~\cite{mishrareentrant} occur as a function of $\lambda$. This type of signature is similar to the situation discussed in Ref.~\cite{mishrareentrant}. However, the interesting feature here is that for a range of $\lambda$ ($1 < \lambda \lesssim 2$), the system undergoes a series of localization transitions as a function of $\Delta$. In the following we discuss the phase diagram of Fig.~\ref{fig:fig1}(a) in detail. 

The signatures of the localization transition can be discerned from the participation ratios (PRs)~\cite{paredesreview,review2,mishrareentrant,dassarma2017} which are considered to be the standard diagnostics to study the localization transitions. We compute the inverse participation ratio (IPR) and the normalized participation ratio (NPR) for the $m^{th}$ eigenstate  defined as ${\rm{IPR}}_{m}=\sum_{i=1}^{L} |\psi^{i}_{m}|^{4}$ and ${\rm{NPR}}_{m}=\bigg(L\sum_{i=1}^{L} |\psi^{i}_{m}|^{4}\bigg)^{-1}$ respectively, where $i$ is the site index~\cite{biddle,Ganeshan,paredesreview}. The values of $\rm{IPR_m} = 0 ~(\neq 0)$ and $\rm{NPR_m}\neq 0 ~(=0)$ in the large $L$ limit characterize the extended (localized) states. 
The re-entrant localization behaviour can be understood by directly plotting the IPR as a function of all the eigenenergies and eigenstates of Eq.~\ref{eq:ham} for different values of $\lambda$ as depicted in Fig.~\ref{fig:fig1}(b) and (d) respectively for an exemplary value of $\Delta=1.8$. While the regions with dark blue (red) color for all the states indicate the extended (localized) phases at weak (strong) values of $\lambda$, two intermediate phases in the range $0.4 \lesssim \lambda \lesssim 0.8$ and $1.4 \lesssim \lambda \lesssim 1.6$ indicate the presence of both extended ($\rm{NPR}\neq 0$) and localized ($\rm{IPR}\neq 0$) states. Based on the behaviour of the PRs of the states we compute the phase diagram shown in Fig.~\ref{fig:fig1}(a) by plotting the quantity $\eta =\log_{10}[\langle \rm IPR\rangle \times \langle \rm NPR\rangle]$~\cite{dassarma2017,mishrareentrant} in the $\Delta - \lambda$ plane. Here $\langle~\cdot~ \rangle$ stands for the average over the entire spectrum. For the extended or localized phase since either $\langle \rm IPR\rangle$ or $\langle \rm NPR\rangle$ has $1/L$ dependence on the system size, $\eta < -\log_{10} L$ i.e. $\eta<-4$ as $L\sim 10^4$ in our case but when both of them are finite, i.e. of $\mathcal{O}(1)$, we get $-4<\eta<-1$. The phase diagram clearly exhibits two localization transitions (re-entrant localization) as a function of $\lambda$ for several values of $\Delta$. However, for some fixed values of $\lambda$, the localization transition is robust as a function of $\Delta$. For smaller $\lambda$, there exists only one localization transition at larger $\Delta$. With increase in $\lambda$ the system exhibits multiple localization transitions as a function of $\Delta$. Interestingly, the number of localization transitions increases up to $\lambda \lesssim 2 = \lambda_{AA}$ where four localization transitions can be seen. 

To quantify these localization transitions as a function of $\Delta$, we plot the extrapolated values of average PRs as a function of $\Delta$ in Fig.~\ref{fig:fig1}(c) for a cut through the phase diagram of Fig.~\ref{fig:fig1}(a) at $\lambda=1.5$. Initially, when $\lambda=1.5$ and $\Delta=0$ (AA limit), the system is in the extended phase indicated by $\langle \rm{IPR}\rangle=0$ and $\langle \rm{NPR}\rangle \neq 0$. With increase in $\Delta$, the system enters into different phases. The behaviour of $\langle \rm{IPR}\rangle$ together with $\langle \rm{NPR}\rangle$ clearly indicates transitions to the localized phases at three different critical values of $\Delta$ such as $\Delta_c \sim 0.7,~1.2,~1.9$ and three intermediate phases emerge (shaded regions).  
These localization transitions are further complemented by inspecting the behaviour of other parameters of interest such as the Shannon entropy~\cite{shanonentropy0,shanonentropy1,Sengupta_2021} and the gap ratio~\cite{subrotoreview,ray_2015}. The Shannon entropy is defined from the single particle states as $S_m=-\sum_i|\psi_m^i|^2\ln(|\psi_m^i|^2)$ which vanishes for the localized states due to participation from a single site only and approaches its maximum value $\ln(L)$ for the extended states where the wave amplitude is finite for all lattice sites~\cite{Sengupta_2021}. On the other hand the average value of the gap ratio $r_m=\frac{min(\delta_m,\delta_{m+1})}{max(\delta_m,\delta_{m+1})}$ where $\delta_m=E_m - E_{m-1}$ is the difference between two adjacent energies calculated using PBC, is $0~(0.386)$ in the extended (localized) phases~\cite{subrotoreview,ray_2015}. Here, $E_m$ are the eigenenergies arranged in increasing order. We plot  $\langle S\rangle/\ln(L)$ and $\langle r \rangle$ as a function of $\Delta$ in Fig.~\ref{fig:fig1}(c). The Shannon entropy clearly vanishes in the localized phases and remains finite in both the extended and the intermediate phases as expected. On the other hand, $\langle r \rangle$ approaches its maximum value (as predicted by the Poissonian statistics) in the localized phases and decreases in the intermediate phases and becomes vanishingly small in the extended phase. All these quantities together confirm the multiple localization transitions of Fig.~\ref{fig:fig1}(a). 

\begin{figure}[t]
\centering
{\includegraphics[width=3.42in]{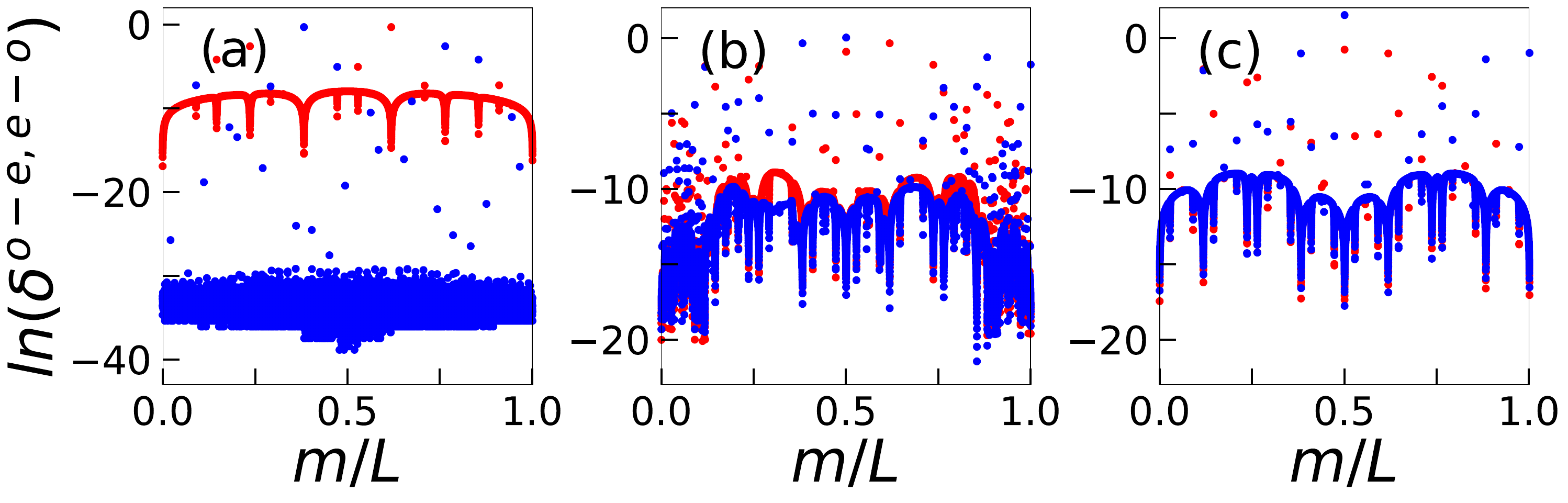}}
\caption{Even-odd $\delta^{e-o}$ (red) and odd-even $\delta^{o-e}$ (blue) for $\lambda=0.75$ and different values of $\Delta$. For (a) $\Delta=0.0$, (b) $\Delta=0.9$ and (c) $\Delta=3.0$ the system lies in the extended, intermediate and localized phase respectively. The results are obtained for a system of size $L=28657$ and $\beta=17711/28657$ under PBC.}
\label{fig:fig4}
\end{figure}

To further understand the behaviour of $\langle r \rangle$  we compute the even-odd (odd-even) level spacing which is given by $\delta^{e-o}=E_{2m}-E_{2m-1}$ ($\delta^{o-e}=E_{2m+1}-E_{2m}$) \cite{santos_deng_refs, Sengupta_2021} 
that can clearly distinguish between the extended, intermediate and localized phases. In the extended region, due to the presence of doubly degenerate spectrum of the underlying Hamiltonian with no onsite staggered potential, $\delta^{e-o}$ vanishes and hence it is well separated from $\delta^{o-e}$ as shown in Fig.~\ref{fig:fig4} (a). On the other hand, $\delta^{e-o}$ and $\delta^{o-e}$ have no clear distinction between them for the localized states as the degeneracy is lifted (Fig.~\ref{fig:fig4} (c)). However, in the intermediate region (Fig.~\ref{fig:fig4} (b)), we see regions of finite and vanishing gap between $\delta^{e-o}$ \& $\delta^{o-e}$ when the states are extended and localized respectively. Therefore, $\langle r\rangle$ attains zero (maximum) value in the extended (localized) regions as $min(\delta_m,\delta_{m+1})$ vanishes for all $m$. However, in the intermediate region, it lies in between the two extreme values.

\begin{figure}[t]
\centering
{\includegraphics[width=3.4in]{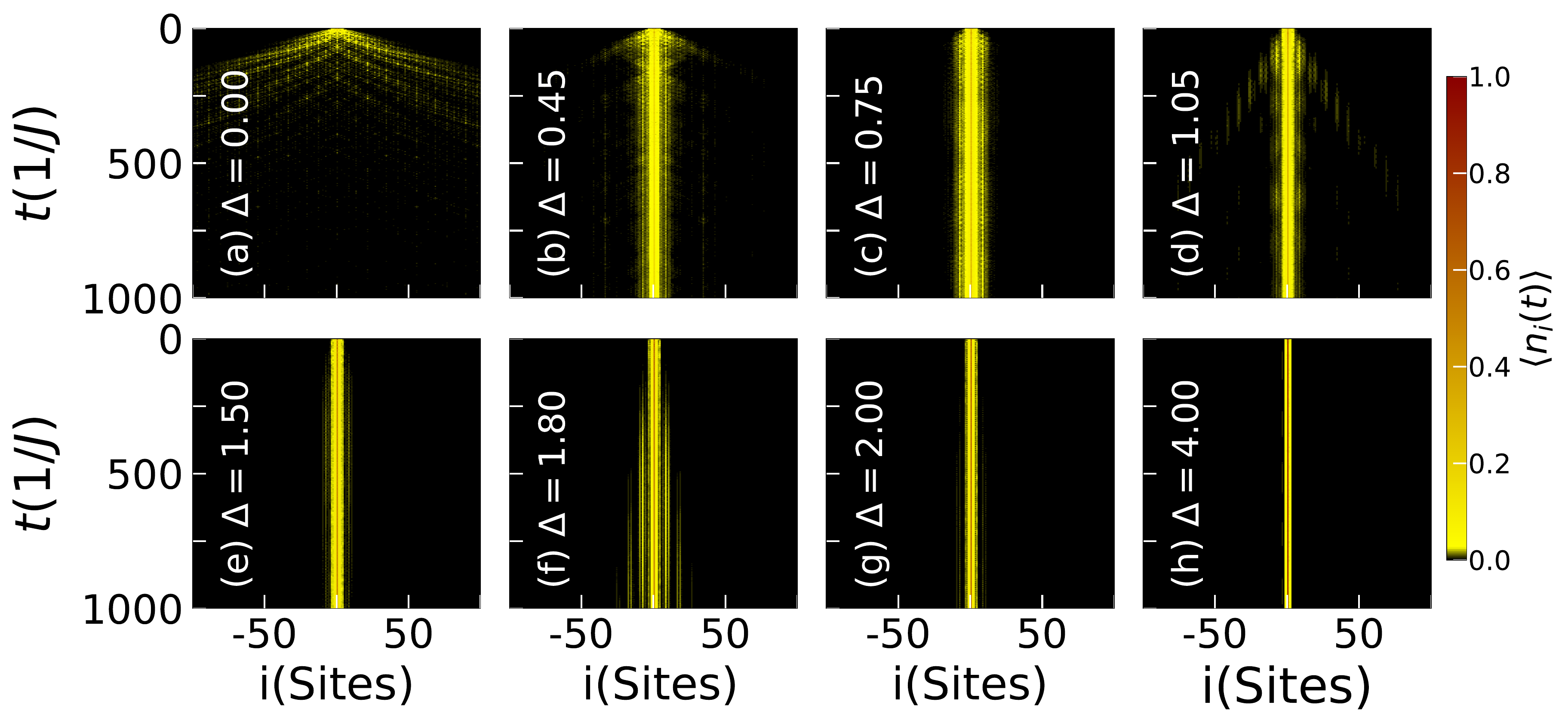}}
\caption{The density distribution during the time evolution of an initial state for different values of $\Delta$ (a-h) at a fixed $\lambda=1.5$ and $L=2584$ with an average over $500$ different values of phase offset $\phi$. Only central $201$ sites are shown for clarity.} 
\label{fig:fig2}
\end{figure}
The important inferences which stem out from our analysis so far are as follows. From Fig.~\ref{fig:fig1}(a) it is clear that in general $\Delta$ favours the localization transition. However, for some values of $\Delta$ the AA limit of the localization transition is unaffected and the (re)localization transition occurs exactly at $\lambda = \lambda_{AA}=2$. On the other hand, as a function of $\Delta$, there occurs only one localization transition in the limit of smaller $\lambda$. However, for larger $\lambda$, the system undergoes multiple localization transitions. While further analysis is necessary to understand the physics behind these intriguing behaviour of multiple localization transitions and the associated intermediate phases, one possible reason for the multiple localization can be as follows. Due to the two competing onsite potentials such as the disorder and the staggered potential, a situation might favour an overlap of two nearest neighbour states turning a localized state extended~\cite{lado2020_nphys}. This kind of re-entrant localization of the individual states happens in different parameter domains depending on the energies of the states. Hence, the system as a whole exhibits multiple localization  transitions.

{\em Expansion dynamics.-} In this part we analyse the multiple localization transitions discussed above in the expansion dynamics. Our analysis is based on the standard quenching protocols of unitary time evolution i.e. $|\Psi(t)\rangle=e^{-iHt}|\Psi(0)\rangle$, with the time independent Hamiltonian $H$ given in Eq.~\ref{eq:ham} and an initial state $|\Psi(0)\rangle$ at $t=0$. For our studies we consider $|\Psi(0)\rangle=|\cdot~\cdot~c_0^\dagger~\cdot~\cdot\rangle$, a state corresponding to a particle located at the center of the lattice with open boundary condition.
\begin{figure}[t]
\centering
{\includegraphics[width=3.5in]{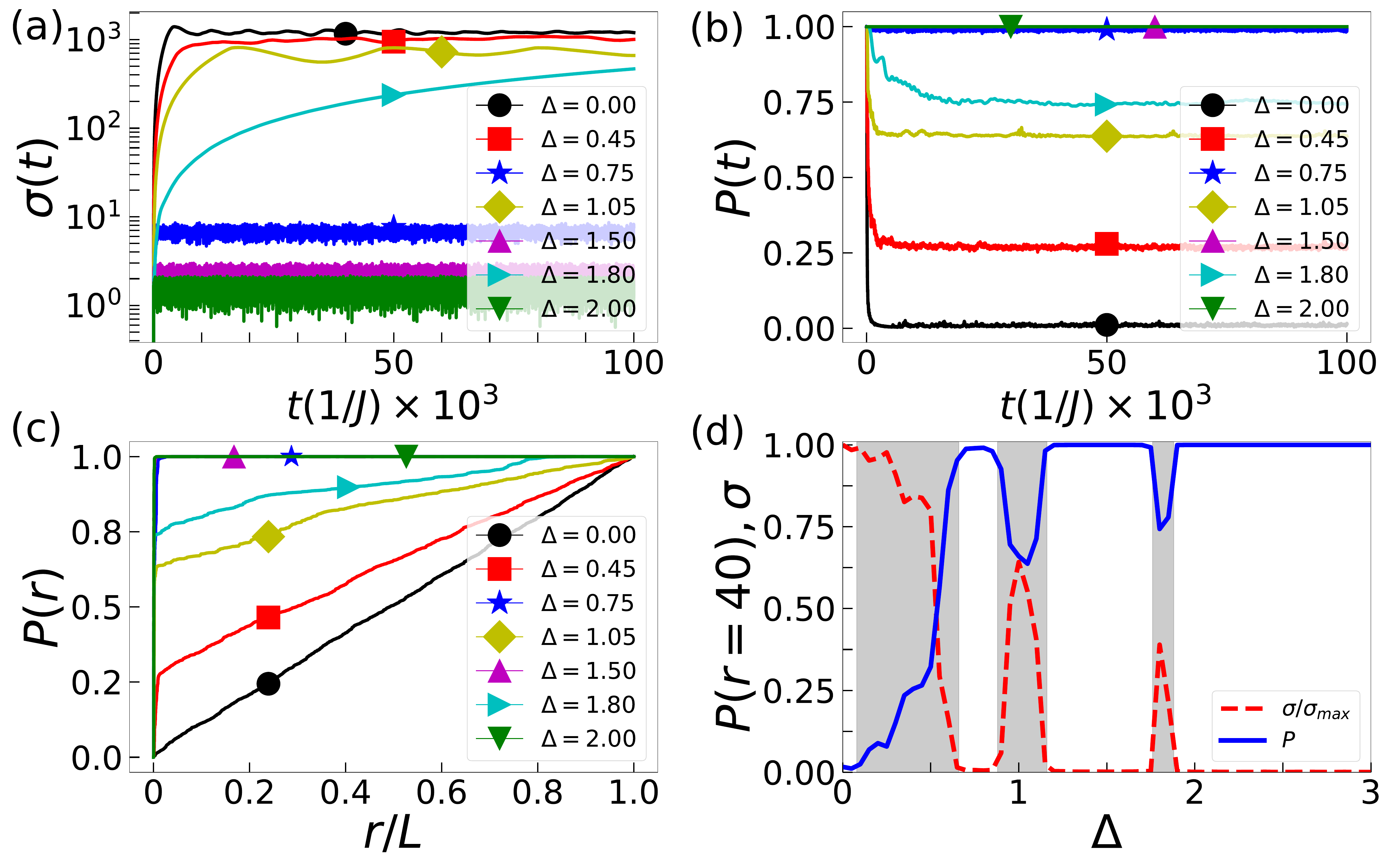}}
\caption{(a) $\sigma(t)$ vs $t$ (b) $P_{r=40}(t)$ vs $t$ for $\Delta=0.0,~0.45,~0.75,~1.05,~1.5,~1.8,~2.0$ and $\lambda=1.5$. (c) $P_r(t=10^5(1/J))$ as a function of $r/L$ shows the multiple localization transition (see text). (d) The time evolved values of $\sigma$ (red dashed) and $P_{r=40}$ (blue solid) to $t=10^5(1/J)$ are plotted as a function of $\Delta$ for $\lambda=1.5$. The shaded regions indicate the intermediate phases. For all the cases a system size of $L=4181$ is considered. }
\label{fig:fig3}
\end{figure}
First of all, we track the real space density evolution $\langle n_i(t)\rangle$ computed using the time evolved state $|\Psi(t)\rangle$ as shown in Fig.~\ref{fig:fig2}(a-h) for different values of $\Delta=0.0,~0.45,~0.75,~1.05,~1.5,~1.8,~2.0,~4.0$ respectively for a cut through the phase diagram at $\lambda=1.5$. While the extended and localized phases are characterized by the fast and no spreading of the densities respectively, the intermediate phases are characterised by a bimodal distribution where part of the density remains localized around the central site and a part expands slowly. The bimodal nature of the density profile is due to the presence of both localized and extended states in the intermediate phase. Although this quantity can be observed in experiments~\cite{Greiner_walk}, to obtain clear signatures of the multiple localization transition we compute the root mean-square displacement defined as $\sigma(t)=\left[\sum_i(i-i_0)^2|\psi_i(t)|^2\right]^{1/2}$, where $i_0$ is the initial position of the particle. This quantity directly corresponds to the expansion of wavepacket and exhibits distinct features in its long time evolution for the extended, intermediate and the localized phases as recently observed in the quantum gas experiments~\cite{lueschen}. 

While a faster (slower) expansion characterizes the extended (intermediate) phases, the localized phases are characterised by no expansion as shown in Fig.~\ref{fig:fig3}(a). 
In the figure, the saturation of  $\sigma(t)$ to different values after a long time evolution~\cite{lueschen,paredesreview} for different $\Delta=0.0,~0.45,~0.75,~1.05,~1.5,~1.8,~2.0,~4.0$ and $\lambda=1.5$ clearly indicates the multiple localization transitions.

 We also compute the survival probability defined as $P_r(t)=\sum_{i=-r/2}^{r/2}|\psi_i(t)|^2$ which is the probability of finding the particle within a range of sites between $-r/2$ to $r/2$ for small $r$~\cite{santos_deng_refs, leasantos,shuchen}. For the quenching to the extended phase, due to the fast expansion of the wavepacket, $P_r(t)$ tends to vanish in the long time evolution. On the other hand for the localized phases, $P_r(t)$ is maximum ($P_r(t)\to 1$) indicating no spreading. However, in the intermediate phases, due to the finite probability of both escaping and remaining within the small range of $r$ the value of $P_r$ decreases but remains finite. To examine this behaviour we plot $P_{r=40}(t)$ as a function of $t$ in Fig.~\ref{fig:fig3}(b) for parameters considered in Fig.~\ref{fig:fig3}(a) and obtain a clear feature of multiple localizations. 
 Additionally, the multiple localization transitions can also be seen by looking at the behaviour of $P_r$ as a function of $r/L$ for different values of $\Delta$ after the long time evolution (Fig.~\ref{fig:fig3}(c)). As in the extended phase ($\Delta=0$), the probability of finding the particle at all sites is equal to $1/L$, the $P_r$ varies as $r/L$ depicting an almost straight line. For the localized phases, $P_r \to 1$ within a short range of $r/L$ as expected. However, for the intermediate phases, $P_r \to 1$ at different values of $r/L$ for a given time $t$. The saturation of $P_r$ is dependent upon the localization length; larger the localization length, $P_r \to 1$ at larger $r/L$ value and vice versa. As $\Delta$ increases, the multiple localization features can be clearly seen as the position of $P_r \to 1$ shifts between the extended and localized limits.  
 
 Finally to clearly see the signatures of the multiple localization transitions from the expansion dynamics, we plot the saturated values of $\sigma(t=10^5(1/J))$ along with $P_r(t=10^5(1/J))$ as a function $\Delta$ for $\lambda=1.5$ and $L=4181$. The variation of $P_r$ along with $\sigma$ clearly shows three localization transitions and three intermediate phases (shaded regions in Fig.~\ref{fig:fig3}(d)) as already shown in Fig.~\ref{fig:fig1}(c).

{\em Experimental scheme.-} In this part we provide a realistic setup to observe the  multiple localization transitions predicted above in optical lattice experiments. It should be noted that the lattice model considered here has already been used as a tool to study the localization and topological physics~\cite{lueschen,bordia,schreiber,Takahashi} in the quantum gas experiments. For the observation of the localization phenomena, a lattice with onsite staggered potential (superlattice) was used to create an initial state for the dynamical evolution and was not part of the final quasiperiodic system. However, in our case the superlattice potential is itself a parameter of the Hamiltonian and moreover, our predictions are based on the dynamics of a single particle initial state. Therefore, we propose a different method where the single particle initial state can be created and the dynamics can be observed. 

Following the prescriptions given in Ref.~\cite{lueschen,Takahashi}, at first a square lattice can be created with superlattice (normal lattice) along the $x$- ($y$-) directions using the equation 
\begin{eqnarray}
 V(x,y)&=&V_P \cos^2\left(\frac{2\pi x}{d}\right)+V_L \cos^2\left(\frac{\pi x}{d} - \varphi\right)\nonumber\\
 &+&  V_P \cos^2\left(\frac{2\pi y}{d}\right) \nonumber
\end{eqnarray}
where $V_P$ and $V_L$ are the depths of the primary and the long lattices respectively, $d$ is the lattice constant and $\varphi$ is the phase difference between them. By choosing $V_P >> V_L$ and $\varphi=\pi/3$ a superlattice relevant for our studies can be created along the $x$-direction whereas the lattice remains uniform along the $y$-direction. When loaded with ultracold atoms (spin-polarized fermions or hardcore bosons), an initial stripe phase can be formed where atoms occupy the rows of deep lattice sites along the $y-$ direction. Further, the atoms can be selectively removed from all the rows except the central row ~\cite{Greiner_walk} leading to a situation where each individual tube contains only one atom localized at the center.  Now by superimposing a disorder lattice of the form $V_D \cos^2(\frac{\pi x}{d_D})$ along the $x$-direction, the desired initial state can be achieved which can be deep in the localized phase of the phase diagram shown in Fig.~\ref{fig:fig1}(a). Following an appropriate quenching protocol, the multiple localization transitions can in principle be observed by measuring the above mentioned quantities.

{\em Conclusions.-} We have studied the localization transitions in a one-dimensional quasiperiodic lattice of AA type with onsite staggered potential. By analysing various physical quantities such as participation ratios, Shannon entropy and gap ratio we have predicted the scenario of multiple localization transitions in the parameter space. The important observation is that for fixed disorder strengths, there exist multiple localization transitions of the system as a function of the staggered potential strength. Interestingly, the number of localization transition as a function of the staggered potential increases with increase in disorder strengths. Apart from this the system undergoes two localization transitions as a function of the disorder strength for different values of the dimerization - a feature similar to Ref.~\cite{mishrareentrant}. All the localization transitions are found to occur through intermediate phases hosting both the extended and localized states. We have further analysed these findings in the context of expansion dynamics and proposed a method for observation of the multiple localization transitions in the quantum gas experiments. 


{\em Note added:} While preparing the manuscript, we became aware of an interesting recent work~\cite{miguel1} following which the mobility edges can be analytically estimated in some limiting situations of the model considered in our studies indicating multiple localization transitions~\cite{miguel2}.

\bibliography{references}

\end{document}